\documentclass[prd,twocolumn,nofootinbib,showpacs,amsmath,amssymb]{revtex4}

\usepackage{graphicx}
\usepackage{dcolumn}
\usepackage{bm}

\begin{document}

\title{Proton (antiproton) elastic scattering at energies from FNAL to LHC\\
in the tripole pomeron-odderon model}

\author{E. Martynov}

\affiliation{%
Bogolyubov Institute
for Theoretical Physics, 03680 Kiev, Ukraine.\\
e-mail: martynov@bitp.kiev.ua
}%

\date{\today}

\begin{abstract}
The model of elastic scattering amplitudes dominated by the  triple (at $t=0$) pomeron pole  suggested earlier  is modified to confront to  existing experimental data on  $pp$ and $\bar pp$ total and differential cross sections at $\sqrt{s}\geq 19$ GeV and $|t|\leq 14.2$ GeV$^{2}$ including the newest TOTEM data.  Predictions for the future TOTEM measurements at 13 and 14 TeV are given.
\end{abstract}

\pacs{13.85.-t, 13.85.Dz, 13.85.Lg, 11.55.Jy}
\maketitle

\section{\label{sec:level1}Introduction}

The TOTEM experiment at LHC provides, in fact, the first measurements of soft pomeron contribution (more exactly of  pomeron
and odderon) because  the contributions of secondary or subasymptotic reggeons are very small at such high energies. It means that  the precise measurement of $pp$ total and differential cross sections \cite{totem1, totem2,totem3} gives a possibility to discriminate the various pomeron models by comparing their predictions with new data. Generally speaking the TOTEM results are going to be a strong restriction for various phenomenological pomeron models \cite{totem1}. Many of them being directly extrapolated to 7 TeV could not describe the recent TOTEM data though they are  in a good agreement with data at $\sqrt{s}\leq $1.96 TeV. Moreover, even basing  on the total cross section measurement at 7 and 8 TeV only one can conclude that $\sigma_{tot}^{pp}$ rises with energy faster than $\ln s$. Therefore  some of the models have to be rejected despite  its excellent agreement  with the data available before LHC epoch. For example, the dipole pomeron model  with trajectory intercept $\alpha(0)=1$ (see  \cite{EMDT} and references therein) is unable to give $\sigma_{tot}^{pp}>94$ mb at $\sqrt{s}=7$ TeV.

The current state of elastic scattering model market is given in \cite{Godizov} and in the comprehensive review \cite{Dremin}. Obviously, all the models must be verified for a consistency with new data. Many of them, if not the all, require some modification to be successfully extrapolated to new data.

 In \cite{AKM} such a procedure was performed for the eikonal \cite{3Pom1,3Pom2} and $U$-matrix \cite{UM} models. It was shown that adding  the second odderon term solves  the problem only partially, a further modification of the model is  necessary. Eikonal and $U$-matrix models are treated as infinite series of  multiple input reggeon exchanges. However another approach to constructing  model is also possible. One can  just  take into account unitarity and analyticity requirements from the beginning  as well as experimental information on the cross sections (e.g. increase of total cross sections) to determine the form of amplitude.

Following this idea in  \cite{EMDT} we have suggested the tripole pomeron-odderon model of elastic $pp$ and $\bar pp$ scattering amplitudes and compared it with data. Amplitudes in the model at very high energies  are dominated by pomeron contribution which corresponds to pair of the branch points colliding at $t=0$ and at  the angular momentum $j=1$ to generate a triple pole. Besides the set of the preasymptotic terms, even and odd, have been incorporated in the model. Structure of leading singularities is different from those of the well-known  the maximal pomeron and odderon model \cite{GLN, AGN, MN}. It was shown that the model \cite{MN} is in a good agreement with the data in a wide interval of energy and momenta transferred, but now we have found that those model without changes  fail to describe the recent TOTEM $d\sigma/dt$ data.

 In the present paper we are focused on our parametrization of the tripole pomeron-odderon model  \cite{EMDT} exploring its simplified form (neglecting the cut contributions unimportant at high energies). This model in ref. \cite{EMDT} was applied at $|t|\leq$ 6 GeV$^{2}$. We add  two additional  terms decreasing at high $|t|$ as $1/|t|^{4}$ in order to extend model for larger $|t|$. Such terms allow us (as it was shown in  \cite{MN}) to describe the $d\sigma/dt$ data  at $|t|\gtrsim$ 6 GeV$^{2}$).

 We remind the important properties of the tripole pomeron-odderon model, argue its modification for high $|t|$, then we present the results of the fit and give predictions for the TOTEM measurements at higher energies and $|t|$.

\section{Tripole pomeron model}\label{sec:Tpar}
 \subsection{The leading high energy terms of the amplitude.}

As it was stressed in \cite{EMDT} the contribution  to the partial amplitude of a triple pole with linear trajectory
\begin{equation}\label{3pole-lin}
\phi(j,t)=\frac{\beta(j,t)}{[j-1-\alpha't]^{3}}
\end{equation}
violates the unitarity inequality $\sigma_{el}(s)\leq \sigma_{tot}(s)$. The correct pomeron singularity dominating the partial wave  is the following
\begin{equation}\label{eq:tripole1}
\varphi_{1{\cal P}}(j,t)=\eta(j)\frac{\beta_{1}(j,t)}{\left
[(j-1)^{2}-kt\right]^{3/2}}.
\end{equation}
The black disk model  holds  the same singularity of the partial amplitude. The imaginary part of black disk amplitude in the impact parameter representation has a step-like form  with  the height 1/2 and  the width equal to $R^{2}(s)\propto \ln^{2} s$. It leads to $\sigma_{tot}(s)\propto \sigma_{el}(s)\propto \ln^{2}(s/s_{0}), \quad s_{0}=1$ GeV$^{2}$.

It seems to be natural  to keep the same structure of singularity for subleading terms changing only their multiplicity. Our choice of subleading terms is the following
\begin{equation}\label{eq:tripole2}
\varphi_{2{\cal P}}(j,t)=\eta(j)\frac{\beta_{2}(j,t)}{\left
[(j-1)^{2}-kt\right]},
\end{equation}
\begin{equation}\label{eq:tripole3}
\varphi_{3{\cal P}}(j,t)=\eta(j)\frac{\beta_{3}(j,t)}{\left
[(j-1)^{2}-kt\right]^{1/2}},
\end{equation}
\begin{equation}
 \eta(j)=\frac{1+e^{-i\pi
j}}{-\sin\pi j}. \nonumber
\end{equation}
Thus  the leading  pomeron contribution to partial amplitude has a form
\begin{equation}\label{eq:j-tripole}
\varphi _{{\cal P}}(j,t)=\varphi_{1{\cal P}}(j,t)+\varphi_{2{\cal P}}(j,t)+\varphi_{3{\cal P}}(j,t).
\end{equation}

Taking into account that
\begin{equation}\label{eq:phi1}
\frac{1}{(\omega^{2}+\omega_{0}^{2})^{3/2}}=\frac{1}{2\omega_{0}}
\int\limits_{0}^{\infty}
dx xe^{-x\omega}J_{1}(\omega_{0}x),
\end{equation}
\begin{equation}\label{eq:phi2}
\frac{1}{\omega^{2}+\omega_{0}^{2}}=\frac{1}{\omega_{0}}\int\limits_{0}^{\infty}
dx e^{-x\omega}\sin(\omega_{0}x),
\end{equation}
\begin{equation}\label{eq:phi3}
\frac{1}{(\omega^{2}+\omega_{0}^{2})^{1/2}}=\int\limits_{0}^{\infty}
dx e^{-x\omega}J_{0}(\omega_{0}x)
\end{equation}
where $J_{0,1}(\omega_{0}x)$ are the Bessel functions, one can write the main pomeron part of amplitude in the $(s,t)$-representation
\begin{equation}\label{eq:tripom-st}
\begin{array}{ll}
&{\cal P}(s,t)=iz\biggl\{{\cal P}_{1}(s,t)+{\cal P}_{2}(s,t)+{\cal P}_{3}(s,t)\biggr \},\\
&{\cal P}_{1}(s,t)=g_{1}^{{\cal P}}v_{1}^{{\cal P}}(t)\xi \frac{2J_{1}(\xi\tau_{+})}{\tau_{+}},\\
&{\cal P}_{2}(s,t)=g_{2}^{{\cal P}} v_{2}^{{\cal P}}(t)\frac{\sin(\xi\tau_{+})}{\tau_{+}},\\
&{\cal P}_{3}(s,t)=g_{3}^{{\cal P}} v_{3}^{{\cal P}}(t)J_{0}(\xi\tau_{+})
\end{array}
\end{equation}
where $g_{i}^{{\cal P}}v_{i}^{{\cal P}}(t)$ are the vertex functions with $ v_{i{\cal P}}(0)=1,\quad \xi =\ln(-iz/z_{0})$, $z$ is defined below by
Exp.(\ref{eq:zvar}), and $\tau_{+}=r_{+}\sqrt{-t/t_{0}}$, $t_{0}=1$
GeV$^{2}$, $g_{i}^{{\cal P}}, r_{+}$ and $z_{0}$ are constants.

Strictly speaking the amplitudes in Regge models depend on $s$  through the $\cos\theta_{t}$, cosine of the scattering angle in $t$-channel, which for
 $pp$ scattering in the center mass system has the form
\begin{equation}\label{eq:costetha}
\cos\theta_{t}= 1+2s/(t-4m_{p}^{2})=(t+2s-4m_{p}^{2})/(t-4m_{p}^{2}).
\end{equation}
 In the considered kinematical region  the  $t$ in the numerator of Eq.~\ref{eq:costetha} can be neglected. However we keep $4m_{p}^{2}$ because for $t=0$ we take into account the data on cross sections  at low energies, $\sqrt{s}\gtrsim$ 5 GeV. Absorbing factor $1/(t-4m_{p}^{2})$ into the vertex functions we define the energy variable as the following
\begin{equation}\label{eq:zvar}
 z=2s-4m_{p}^{2}.
\end{equation}

We propose to keep the same singularity structure for the  leading  terms of crossing-odd part of amplitude. However,  taking into account the fact that there is no visible odderon contribution at $t=0$ we multiply each odderon term by factor $t$
\begin{equation}\label{eq:triodd-st}
\begin{array}{ll}
&{\cal O}(s,t)=zt\biggl \{{\cal O}_{1}(s,t)+{\cal O}_{2}(s,t)+{\cal O}_{3}(s,t)\biggr \},\\
&{\cal O}_{1}(s,t)=g_{1}^{{\cal O}}v_{1}^{{\cal O}}(t)\xi \frac{2J_{1}(\xi \tau_{-})}{\tau_{-}},\\
&{\cal O}_{2}(s,t)=g_{2}^{{\cal O}}v_{2}^{{\cal O}}(t)\frac{\sin(\xi \tau_{-})}{\tau_{-}},\\
&{\cal O}_{3}(s,t)=g_{3}^{{\cal O}}v_{3}^{{\cal O}}(t)J_{0}(\xi \tau_{-})
\end{array}
\end{equation}
where $v_{i}^{{\cal O}}(0)=1, \tau_{-}=r_{-}\sqrt{-t/t_{0}}$ and $g_{i}^{{\cal O}}$ are the constants.

We would like to emphasize  that  similar (but not the same) models for ${\cal P}$ and ${\cal O}$ were considered in \cite{GLN, AGN, MN}. These models have the same dominating even  term  ${\cal P}_{1}(s,t)$ but the different subasymptotic terms  ${\cal P}_{2}(s,t)$ ${\cal P}_{3}(s,t)$. The odderon terms  included in those amplitudes do not vanish at $t=0$. The properties of the models \cite{GLN, AGN} and  their defects  have been discussed in detail in \cite{EMDT}. They were modified in \cite{MN}, defects were eliminated and a good description of data at energies up to 1.8 Tev has been obtained.  However, as was noticed in  \cite{Troshin}   such a maximal odderon term  (${\cal O}_{1}(s,t)/t$ in our notations) gives rise to the contradiction with unitarity in the models where $\sigma_{el}(s)/\sigma_{tot}(s)\to const\leq 1$ at $s\to \infty$. In despite of this fact we refit the model \cite{MN} (without the cuts important only at low energies) at the energies $\sqrt{s}>$ 19 GeV, including TOTEM data, but we are fail to describe the data qualitatively.

\subsection{The subleading reggeons and power-like behaved terms of amplitudes.}

In \cite{EMDT} the described model  was applied to analysis of the $d\sigma/dt$ data at $\sqrt{s}>$ 6 GeV, where not only contributions of the pomeron and secondary reggeons but also their rescatterings (or cuts) are very important.  Besides, the considered momenta  transfer were restricted by $|t|_{max}=6$ GeV$^{2}$. Here we would like to check a principal possibility of the model to describe an interpolation  between  GeV  and TeV energy region. We  investigate  as well which amplitude terms  are important for that and  which form  of vertex functions $gv(t)$ is more suitable  for $d\sigma/dt$ at $|t|\gtrsim 5$ GeV$^{2}$. Thus we consider high energy $pp$ and $\bar pp$  elastic scattering, starting from FNAL energy 19 GeV. We do hope that in this energy interval we can neglect at least all cuts reducing the number of adjustable  parameters. These terms are important at lower energies but their parametrization should be chosen being consistent with good description of  high energy data.

We keep in  amplitudes also  the standard "soft"  pomeron and odderon, simple $j$-poles with linear trajectories $\alpha(t)=1+\alpha't$
\begin{equation}\label{eq:softpom}
P(s,t)=-g_{P}v_{P}(t)(-iz/z_{1})^{1+\alpha'_{P}t}, \quad  z_{1}=1 {\rm GeV}^{2},
\end{equation}
\begin{equation}\label{eq:softodd}
O(s,t)=itg_{O}v_{O}(t)(-iz/z_{1})^{1+\alpha'_{O}t},
\end{equation}
To describe the behaviour of the $\sigma_{tot}(s), \rho)(s), d\sigma/dt(s,t)$ at low energy and low $|t|$  the usual secondary crossing-even and odd reggeons ($f, a_{2}, \omega, \rho, ...$ reggeons with intercepts $\alpha (0)\approx 0.4- 0.7$) have to be included in the amplitudes. However, analyzing $pp$ and $\bar pp$ amplitudes only  it is sufficient to consider at $\sqrt{s}>$ 5 GeV one effective even reggeon and one effective  odd reggeon
\begin{equation}\label{eq:sec-reg}
R_{\pm}(s,t)=\binom{-1}{i}g_{\pm}v_{\pm}(t)(-iz/z_{1})^{\alpha_{\pm}(0)+\alpha_{\pm}'t}.
\end{equation}

A preliminary analysis of the data was performed to understand how a behavior of $d\sigma/dt$ (at $|t|\gtrsim 5$ GeV$^{2}$) can be described. We found that a correct description of these data is provided by the terms behaving as $1/(-t)^{4}$ at large $|t|$ (which are almost independent on $s$). Therefore we add to the amplitudes crossing even and odd terms $Ev(s,t)$ and $Od(s,t)$ with arbitrary phases.  To decrease an influence of these terms on the amplitudes at small $|t|$ we introducing the factor $(-t)$. It would be reasonable to construct them in a more customary (Regge-like) form at small $t$ and power-like behaving at large $t$ . But as we noticed above our aim is, first of all, to check some principal possibilities of the tripole pomeron and odderon models. Therefore to avoid an extra number of parameters  we choose these terms in a simplified form
\begin{equation}\label{eq:power+}
Ev(s,t)=i(-t)z\frac{g_{r+}+ig_{i+}}{(1-t/t_{Ev})^{5}},
\end{equation}
\begin{equation}\label{eq:power-}
Od(s,t)=(-t)z\frac{g_{r-}+ig_{i-}}{(1-t/t_{Od})^{5}}.
\end{equation}
Thus the $pp$ and $\bar pp$ amplitudes in a general case are defined as following
\begin{equation}\label{amplitudes}
\begin{array}{ll}
A^{\bar pp}_{pp}(s,t)&={\cal P}(s,t)+P(s,t)+R_{+}(s,t)+Ev(s,t) \\
&\pm \bigg ({\cal O}(s,t) +O(s,t)+R_{-}(s,t)+Od(s,t)\bigg )
\end{array}
\end{equation}
where ${\cal P}, {\cal O},  P, O, R_{\pm}, Ev, Od$ are given by Eqs. (\ref{eq:tripom-st}, \ref{eq:triodd-st}, \ref{eq:softpom}, \ref{eq:softodd}, \ref{eq:sec-reg}, \ref{eq:power+},   \ref{eq:power-}) correspondingly.

\subsection{Choice of the vertex functions $v(t)$ }
We have considered two options for the vertex functions $v(t)$ in the pomeron (${\cal P}(s,t), P(s,t)$) and odderon ((${\cal O}(s,t), O(s,t)$)) terms.

{\bf Model I}, exponential form
\begin{equation}
\begin{array}{lll}
v_{i}^{{\cal P}}(t)&=\exp(2b_{i}^{{\cal P}}t), \quad v_{i}^{{\cal O}}(t)&=\exp(2b_{i}^{{\cal O}}t), \\
v_{P}(t)&=\exp(2b_{ P}t), \quad v_{O}(t)&=\exp(2b_{O}t).
\end{array}
\end{equation}

{\bf Model II}, power form
\begin{equation}
\begin{array}{lll}
v_{i}^{{\cal P}}(t)&=(1-b_{i}^{{\cal P}}t)^{-4}, \quad v_{i}^{{\cal O}}(t)&=(1-b_{i}^{{\cal O}}t)^{-5}, \\
v_{P}(t)&=(1-b_{i}^{ P}t)^{-4}, \quad v_{O}(t)&=(1-b_{i}^{O}t)^{-5}.
\end{array}
\end{equation}
The behaviour with -5  for odderon vertices is taken because odderon terms of amplitudes (Eqs.(\ref{eq:triodd-st},\ref{eq:softodd})) have been multiplied by the factor $t$.

In both models the  vertices for secondary reggeons $R_{\pm}(s,t)$ are chosen in an exponential form
\begin{equation}
v_{\pm}(t)=\exp(2b_{\pm}t).
\end{equation}
$R_{\pm}(s,t)$ are negligible at high $|t|$ because of large slopes of their trajectories.  However these terms of amplitude become important at small $|t|$ where exponential vertices are reasonable.

The following normalization of $p(\bar p)p\rightarrow p(\bar p)p$ amplitude is used
\begin{equation}\label{eq:norm}
\sigma_t=\frac{k}{F_{0}}\Im mA(s,0), \qquad
\frac{d\sigma}{dt}=\frac{k}{F}|A(s,t)|^{2}
\end{equation}
where
\begin{equation}\label{eq:norma}
\begin{array}{ll}
F_{0}&=\sqrt{(s-2m_{p}^2)^2-4m_p^4}=2p_p^{lab}\sqrt{s}, \quad F=16\pi F_{0}^{2},\\
k&=0.3893797{ \rm mb}\cdot{\rm GeV}^{-2}
\end{array}
\end{equation}
and $p_p^{lab}$ is the momentum of initial proton (antiproton) in laboratory system of another proton.
The amplitudes and couplings $g$ are dimensionless with this normalization .

\section{Confronting the models to the data}
\subsection{The data}

The constructed models  were compared with the $pp$ and $\bar pp$ data on $\sigma_{tot}(s), \rho(s))$ and $d\sigma(s,t)/dt$ in  the following  region of $s$ and $t$
\begin{equation}\label{eq:st-region}
\begin{array}{lll}
{\rm for} \quad \sigma_{tot}(s), \rho(s) \quad {\rm at} &\quad 5 {\rm GeV}&  \leq \sqrt{s}\leq 8 {\rm TeV}, \\
{\rm for} \quad d\sigma(s,t)/dt \quad {\rm at} &\quad 19 {\rm GeV}&< \sqrt{s}\leq 7 {\rm TeV}\\
{\rm and}  &\quad 0.01  {\rm GeV}^{2}& \leq |t|\leq 14.2 {\rm GeV}^{2}
\end{array}
\end{equation}
The cosmic ray data on the $pp$ total cross sections were not included in the fit procedure.

The data set we used for  adjusting  the model parameters has been proposed in \cite{CLM}, where a coherent set of all existing data for $4\le \sqrt{s} \le 1800$ GeV and $0\le\vert t\vert\le 14.2$ GeV$^2$ has been built.  A detailed study of the systematic errors of the data  from more than 260 subsets of the data from more than 80 experimental papers have performed (original data and the corresponding references  can be found  in the HEP DATA system \cite{HEPdata,DurhamDB}). The corrected data  are collected and written in a common format. We suggest to use  this data set as a \textit{standard data set}. The latest (with some corrections) updated version of it  including TOTEM \cite{totem1,totem2,totem3} and D0 \cite{D0} data is available online \cite{CLMdata}.

The set used for the given analysis contains 2384 points In the region described above (numbers of points for measured quantities are given in the Table \ref{tab:exppoints}). 31 points from the 3 groups, 8 points at $\sqrt{s}=$26.946 GeV, 11 points at 30.7 GeV and 12 points at 53.018 GeV  only for  $d\sigma_{pp}/dt$ were excluded from the all data presented in \cite{DurhamDB} because these groups are strongly deviated from the rest data points and can slightly distort the fit.

\begin{table}[h!]
\caption{Number of experimental points used for the fitting}
\label{tab:exppoints}
\begin{tabular}{c|rrr}
  &$\quad \sigma_{tot}$  &$ \quad  \rho$   & $ \quad d\sigma/dt$  \\
\hline
$pp$  & 107& 64& 1633\\
$\bar pp$ & 59 & 11 & 510 \\
\end{tabular}
\end{table}

\subsection{Results of the fit}
At the first step we have considered the Models I and II without contribution of $P(s,t)$ and $O(s,t)$ in order to see how  these terms are important. We found out  that such simple models quite well describe the data. The values of $\chi^{2}/{\rm dof}$ (dof$=N_{exp. points}-N_{parameters}$) for all the considered  parametrizations are given in the Table~\ref{tab:chi2}.  The different numbers of parameters in the Models I and II for the case $P(s,t), O(s,t)\neq 0$ are obtained because some of the parameters $b$ are fixed at the low limit $b=0$. The full sets of parameters and corresponding errors for these cases are presented in the Table~\ref{tab:parsers}.

\begin{table}[h!]
\begin{center}
\caption{ The values of $\chi^{2}$ and number of parameters in the considered models}
\label{tab:chi2}
\begin{tabular}{|c|c|c|}
\hline
                       &    \multicolumn{2}{c|}{$\chi^{2}$/dof ($N_{parameters}$)}       \\
\cline{2-3}
                       & {\bf Model  I},& {\bf Model II},  \\
                       & exponential vertices &  power vertices\\
\hline
$P(s,t)=O(s,t)=0$ &     1.779  (26)    &    1.636 (26)      \\
\hline
$P(s,t),O(s,t)\neq 0 $    &  1.402  (31)       &    1.371 (32)      \\
\hline
\end{tabular}
\end{center}
\end{table}

\begin{table}[h!]
  \centering
  \caption{Parameters of the models I and II obtained by fitting to the data.
  Parameters $\alpha'$ and $b$  are given in GeV$^{-2}$,  $z_{0}$ is given in GeV$^{2}$, other parameters are dimensionless. Errors are taken from the MINUIT output.}
  \label{tab:parsers}
  \medskip
{\small 
 \begin{tabular}{|l|c|c||c|c|}
\hline
 & \multicolumn{2}{c||}{\bf Model I} &
\multicolumn{2}{c|}{\bf Model Il}\\
 & \multicolumn{2}{c||}{exponential vertices} &
\multicolumn{2}{c|}{power vertices}\\
\hline
 parameter & value & error &  value & error\\
\hline  $z_{0}$           &   48.438   & 4.127  & 25.438 & 0.644     \\
\hline  $g_{1}^{{\cal P}}$&   0.314   & 0.005  & 0.358 & 0.002      \\
\hline  $g_{2}^{{\cal P}}$&   1.398   & 0.059  &-0.007 & 0.001     \\
\hline  $g_{3}^{{\cal P}}$&   21.215   & 1.164  & 1.843 & 0.22     \\
\hline  $g_{1}^{{\cal O}}$&  -1.018   & 0.062  &-0.174 & 0.004    \\
\hline  $g_{2}^{{\cal O}}$&   0.656   & 0.104 &  1.643 & 0.020    \\
\hline  $g_{3}^{{\cal O}}$&  -11302.3   & 6404.7  &-491.034 &3.079     \\
\hline  $r_{+}$           &   0.281   & 0.003  & 0.372 & 0.001    \\
\hline  $r_{-}$           &   0.682   & 0.009  & 0.201 & 0.001    \\
\hline  $b_{1}^{{\cal P}}$&   3.705   & 0.054  & 0.687 & 0.004    \\
\hline  $b_{2}^{{\cal P}}$&   1.387   & 0.015  & 4.580 & 0.265   \\
\hline  $b_{3}^{{\cal P}}$&   3.035   & 0.642  & 0.180 & 0.022    \\
\hline  $b_{1}^{{\cal O}}$&   2.471   & 0.044  & 1.163 & 0.008     \\
\hline  $b_{2}^{{\cal O}}$&   1.569   & 0.054  & 1.151 & 0.005     \\
\hline  $b_{3}^{{\cal O}}$&   151.69   & 28.98  & 0.419 & 0.0001    \\
\hline  $\alpha_{+}(0)$   &   0.670   & 0.011  & 0.585 & 0.004      \\
\hline  $\alpha_{-}(0)$   &   0.4668   & 0.012  & 0.461 & 0.007   \\
\hline  $\alpha'_{+}$     &   0.84   & fixed  & 0.84 & fixed      \\
\hline  $\alpha'{-}$      &   0.93   & fixed  & 0.93 & fixed   \\
\hline  $g_{+}$           &   74.218   & 2.506  & 74.889 & 1.164      \\
\hline  $g_{-}$           &   58.270   & 3.682  & 60.033 & 2.247       \\
\hline  $b_{+}$           &   0.0  & fixed   & 7.478 & 0.536       \\
\hline  $b_{-}$           &  3.834   & 1.058  & 99.607 & 30.374       \\
\hline $\alpha'_{P}$      &   0.414   & 0.010  & 0.368 & 0.002       \\
\hline $\alpha'_{O}$      &   0.158   & 0.003  & 0.149 & 0.001   \\
\hline $g_{P}$            &   16.772   & 0.923  & 41.892 & 0.192    \\
\hline $b_{P}$            &   0.455   & 0.037  & 1.444 & 0.012    \\
\hline $g_{O}$            &  -.0.063   & 0.005  & 226.08 & 1.753      \\
\hline $b_{O}$            &   0.0   & fixed  & 0.706 & 0.002      \\
\hline $g_{r+}$           &  -39.575   & 1.481  &-6.578 & 0.078    \\
\hline $g_{i+}$           &  -10.945   & 0.477  & 0.132 &  0.072   \\
\hline $t_{Ev}$           &   0.503   & 0.006  & 1.013 & 0.005   \\
\hline $g_{r-}$           &   76.320   & 5.872  &-0.007 & 0.002      \\
\hline $g_{i-}$           &   0.4083   & 0.541  &-0.058 &  0.004     \\
\hline $t_{Od}$           &   0.083   & 0.010  & 4.205 &  0.116    \\
\hline
\end{tabular}
}
\end{table}

\begin{figure}[h!]
\begin{center}
\includegraphics[scale=0.45]{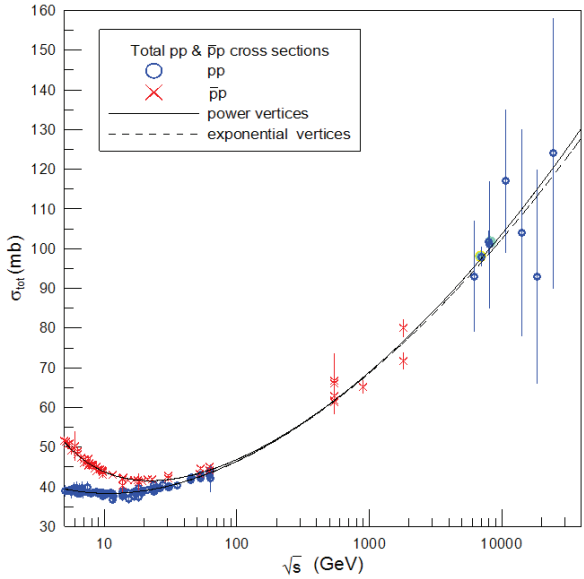}
\caption{$pp$ and $\bar pp$ total cross sections}
\label{fig:sigtot}
\end{center}
\end{figure}

\begin{figure}[h!]
\begin{center}
\includegraphics[scale=0.45]{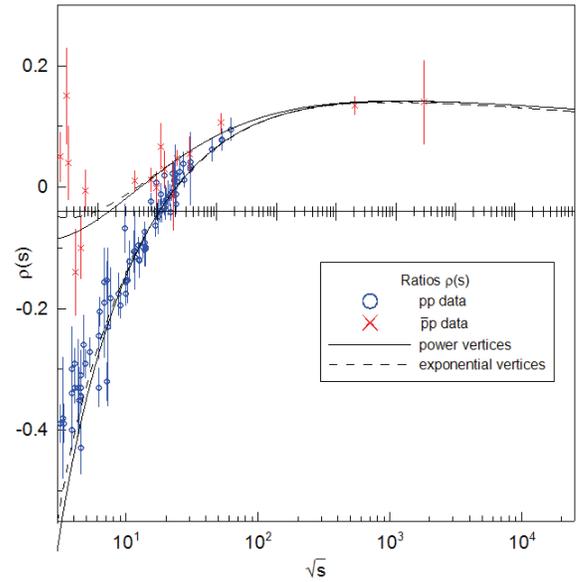}
\caption{Real to Imaginary ratios of the forward scattering $pp$ and $\bar pp$ amplitudes}
\label{fig:rho}
\end{center}
\end{figure}

\begin{figure}[h!]
\begin{center}
\includegraphics[scale=0.45]{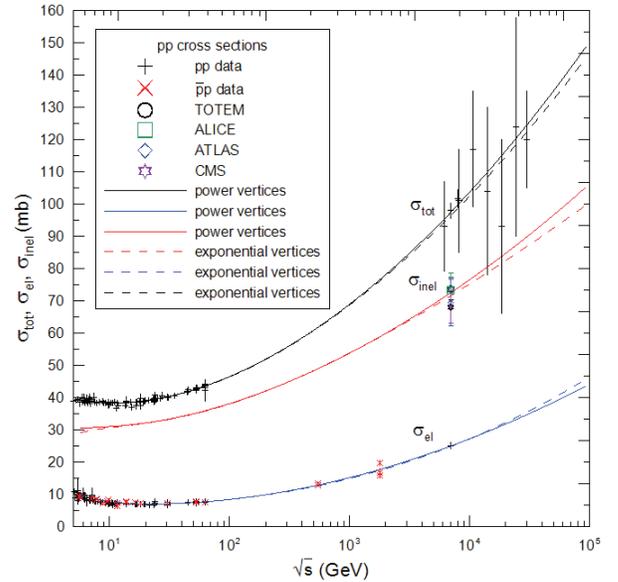}
\caption{Total, elastic and inelastic $pp$ cross sections}
\label{fig:sigint}
\end{center}
\end{figure}

\begin{figure}[h!]
\begin{center}
\includegraphics[scale=0.45]{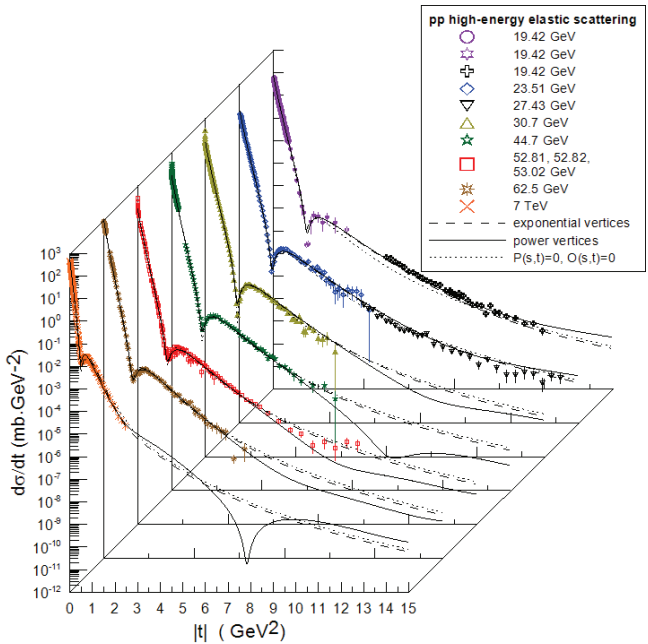}
\caption{Differential cross sections of $pp$ elastic scattering}
\label{fig:pp-dsdt}
\end{center}
\end{figure}

\begin{figure}[h!]
\begin{center}
\includegraphics[scale=0.45]{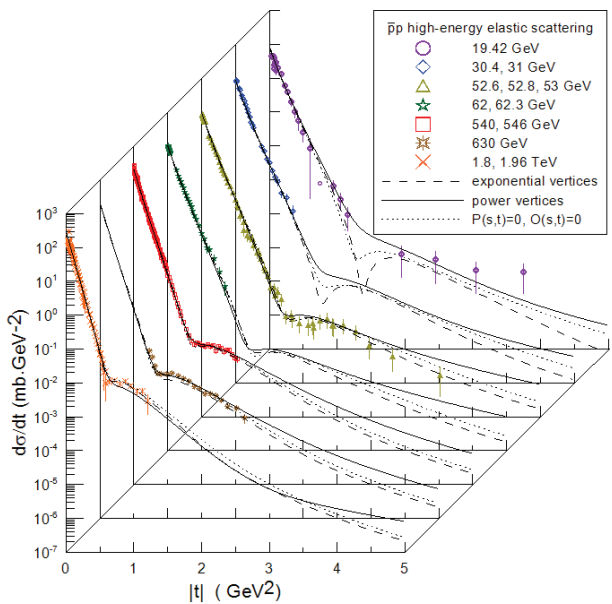}
\caption{Differential cross sections of $\bar pp$ elastic scattering}
\label{fig:pap-dsdt}
\end{center}
\end{figure}

As one can see from the Figs.~\ref{fig:sigtot}-\ref{fig:pap-dsdt} and from the Table~\ref{tab:chi2} all models, even the simplest one with $P(s,t)=O(s,t)=0$, describe quite well the experimental data. The theoretical curves in the models differ in some details but nevertheless they are globally in agreement with experimental data. Moreover, one can obtain very similar description of the data  with the following vertices
\begin{equation}
v(t)=\left \{
\begin{array}{ll}
&\exp(2m_{\pi}-\sqrt{4m_{\pi}^{2}-t}) \quad \text{for even terms},\\
&\exp(3m_{\pi}-\sqrt{9m_{\pi}^{2}-t}) \quad \text{for odd terms}.
\end{array}
\right .
\end{equation}
This first analysis gives a ground for  three important inferences.
\begin{itemize}
\item There are no any unusual or unexpected phenomena in new TOTEM data. They correspond to the standard Regge-like behaviour at low and intermediate $|t|$.
\item  The data at relatively low energies but at highest measured $|t|$ require in amplitude the terms decreasing like $1/(-t)^{4}$. It can be an indication of a hard scattering. However, taking
into account that slow decreasing with $t$ is obtained in the eikonal and $U$-matrix models \cite{AKM}, one can think that a power-like behaviour may be imitated by series of rescatterings or multi-reggeon exchanges.
\item We believe that successful description  of the data in the  considered models results from the well tuned structure of pomeron and odderon singularities
(Eqs.~(\ref{eq:tripom-st},\ref{eq:triodd-st})) rather  than from a choice of vertex functions.
\end{itemize}
The theoretical curves for  $\sigma_{tot}(s), \rho(s)$ and integrated elastic cross sections, $\sigma_{el}(s)$, as well as for inelastic cross section, $\sigma_{inel}(s)$,  are shown in Figs.~\ref{fig:sigtot},\ref{fig:rho},\ref{fig:sigint}. The values of the total cross sections obtained at 7 and 8 TeV are little bit less than the TOTEM data but they are within the measured errors. The both models predictions  for higher LHC energies are given in the Tables~\ref{tab:lhc-predict-e}, \ref{tab:lhc-predict-p}

\begin{table}[h!]
  \centering
  \caption{Cross sections and ratio of real to imaginary part of the forward scattering $pp$ amplitude in the Model I (with $P(s,t), O(s,t)\neq 0$ and exponential vertices)}
  \label{tab:lhc-predict-e}
  \medskip
{\small
 \begin{tabular}{ccccc}
\hline
$\sqrt{s}\,  (TeV)$  & $\sigma_{tot}$ (mb) & $\sigma_{el}$ (mb) &$\sigma_{inel}$ (mb)  &$\rho$  \\
\hline
7    & 96.46 & 24.87 &71.59       &0.132  \\
\hline
 8    & 98.65 & 25.72 &72.93       &0.132  \\
\hline
13   & 106.93  & 29.04  &  77.89   &0.129  \\
\hline
14  & 108.23  & 29.57  & 78.67     &0.128  \\
\hline
\end{tabular}
}
\end{table}

\begin{table}[h!]
  \centering
  \caption{Cross sections and ratio of real to imaginary part of the forward scattering $pp$ amplitude in the Model II (with $P(s,t), O(s,t)\neq 0$ and power vertices)}
  \label{tab:lhc-predict-p}
  \medskip
{\small
 \begin{tabular}{ccccc}
\hline
$\sqrt{s}\,  (TeV)$  & $\sigma_{tot}$ (mb) & $\sigma_{el}$ (mb) &$\sigma_{inel}$ (mb)  &$\rho$  \\
\hline
7    & 97.48 & 24.97 &72.51        &0.136  \\
\hline
 8    & 99.76 & 25.77 &73.98       &0.135  \\
\hline
13   & 108.37  & 28.85  &  79.52   &0.132  \\
\hline
14  & 109.73  & 29.33  & 80.39     &0.132  \\
\hline
\end{tabular}
}
\end{table}

Description of the differential cross section in the models is demonstrated in the Figs.~\ref{fig:pp-dsdt},~\ref{fig:pap-dsdt}. For the simplest model, i.e. at $P(s,t)=O(s,t)=0$, we have shown curves only for model with exponential vertices in order to avoid a meshing figure. From the Table~\ref{tab:chi2} one can see that quality of description in the both models is almost the same. The TOTEM $d\sigma/dt$ data are described with $\chi^{2}/N_{p}=0.47$ and $\chi^{2/}N_{p}=0.46$ in the Models I and II (with $P(s,t),O(s,t)\neq 0$), correspondingly.

\begin{figure}[h!]
\begin{center}
\includegraphics[scale=0.45]{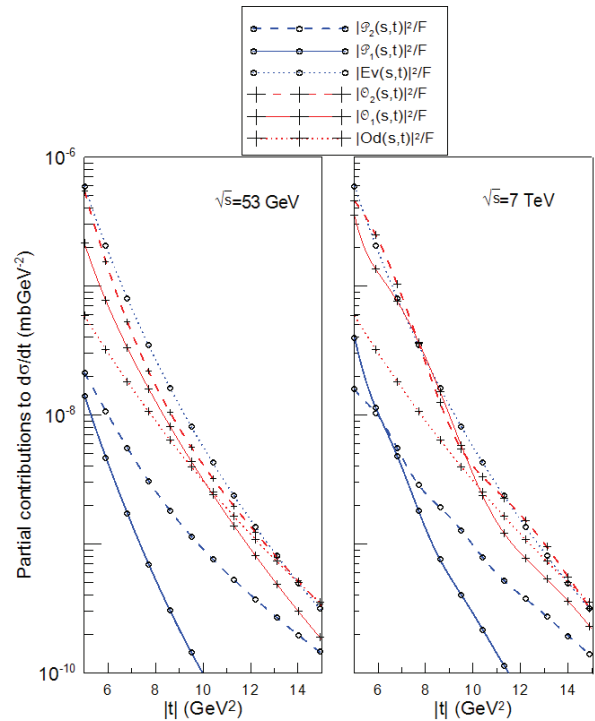}
\caption{Even and odd components of $pp$ elastic scattering amplitudes at 53 GeV and 7 TeV (calculated in the Model II). Factor $F$ is defined in Eq.~\ref{eq:norma}. }
\label{fig:evenodd}
\end{center}
\end{figure}

\begin{figure}[h!]
\includegraphics[scale=0.45]{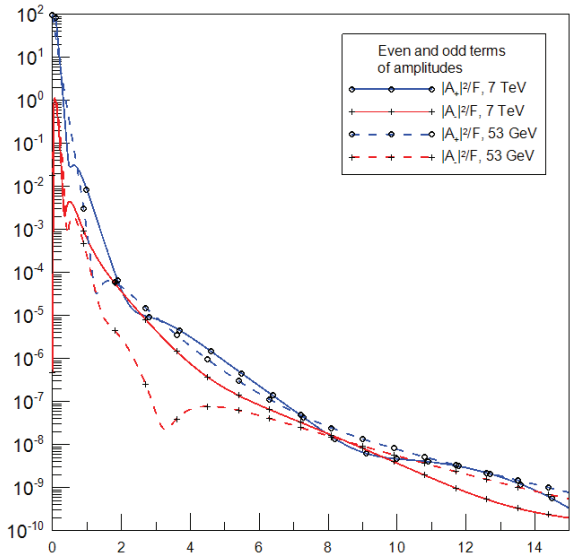}
\caption{Even and odd cumulative contributions to $d\sigma/dt$\\ at 53 GeV and 7 TeV (calculated in the Model II). Factor $F$ is defined in Eq.~\ref{eq:norma}.}
\label{fig:isrlhcpart}
\end{figure}

We would like to stress that in despite of a widespread opinion that at high energy and high momentum transfers an odderon contribution is dominating we found that in the considered models it is not the case. One can see in Figs.~\ref{fig:evenodd},~\ref{fig:isrlhcpart} that  the dominating partial even and odd components at large $|t|$ have comparable  values (at least for $|t|<$15 GeV$^{2}$).  However, Fig.~\ref{fig:isrlhcpart} shows that  cumulative even contribution to  $ds/dt$  at  $|t|$  outside of the dip positions (see below) is larger of the odd one. The cumulative even contribution in this region at 7 TeV is few times larger  than odd  one.   At the same time a role of odd contributions is important at low and intermediate $t$-values  and, especially, in the regions of  dips.

\begin{figure} [h!]
\begin{center}
\includegraphics[scale=0.5]{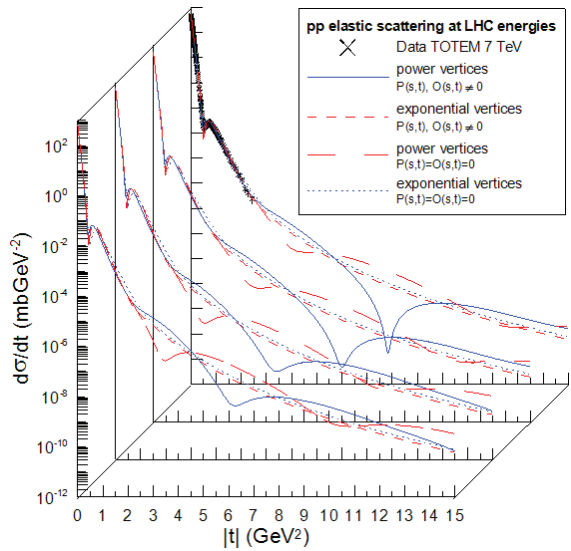}
\caption{Predictions of the models for $pp$ differential cross sections at the LHC energies}
\label{fig:lhc-dsdt}
\end{center}
\end{figure}

In Fig.~\ref{fig:lhc-dsdt} we give the predictions of the considered models for the TOTEM experiment  at higher transferred momenta and higher LHC energies. The most interesting point is an existing/absence  of the second dip in $d\sigma/dt$. Model with exponential formfactors  leads to smooth behaviour with no the dips. Whereas, models with power vertices predict a  dip structure in $d\sigma/dt$. If $P(s,t)=O(s,t)=0$ then the shoulder at $|t|\approx 4$ GeV$^{2}$ and $\sqrt{s}=$ 7 TeV is transformed into a dip moving to $|t|\approx 3$ GeV$^{2}$ at 14 TeV. Moreover, in this model the third dip is visible at the energy 14 TeV near $|t|\sim$ 10 GeV$^{2}$. If $P(s,t),O(s,t)\neq 0$ then well pronounced dip develops  at $|t|\approx$ 8 GeV$^{2}$ moving to $|t|\approx$ 6 GeV$^{2}$ when energy is increasing from 7 to 14 GeV.  Accurate measurements at such $|t|$ would be the excellent test for the considered models.

\section{Conclusion}
Analysis performed above demonstrated a high credibility of the tripole pomeron-odderon model developed in \cite{EMDT}.  We have made its  minor  improvement only by adding two terms, crossing even ($E(s,t)$) and odd ($Od(s,t)$),  in order to apply the model to high $t$.  We have considered  two choices of $t$-dependence in vertex functions and found out in all the cases a good description of the data in wide region of $s$ and $t$ even in the simplified versions of the model without the  standard simple pole pomeron and odderon contributions. Thus, it allow us to conclude that the model is quite stable under variations of the preasymptotic components and the form of vertex functions.  Apparently such the model stability is stipulated by the well tuned structure of the leading pomeron ${\cal P}$ and odderon ${\cal O}$ singularities.

The whole bulk of high energy data are described in a framework of the traditional Regge approach.  In  our opinion  the new data in TeV energy region do not show any indication of new unusual phenomena.

We have predicted the values of the total, elastic and inelastic  cross sections, as well as  the ratio $\rho(s)$ and differential cross sections  at higher LHC energies.
The  amizing prediction is made for large $|t|$-region at the LHC energies. Model with an exponential $t$-dependence of vertices leads to a smooth behaviour of differential $pp$ cross sections while in the model with power vertices a dip structure moving with energy is generated at large $|t|$. We hope that future measurements of TOTEM Collaboration will allow us to discriminate within possibilities for the developed model.

I'd like to thank my colleagues J.R. Cudell, V. Petrov, A. Prokudin, J.P. Revol, S. Troshin and G. Zinovjev  for numerous fruitful discussions of the Regge approaches and high-energy models, for reading the manuscript  and valuable remarks.

The work is supported partially by the Physics and Astronomy Department of National Academy of Sciences of Ukraine (Agreement 2013).

\end{document}